\documentclass[fleqn,usenatbib]{mnras}

\usepackage{newtxtext,newtxmath}

\usepackage[T1]{fontenc}

\DeclareRobustCommand{\VAN}[3]{#2}
\let\VANthebibliography\thebibliography
\def\thebibliography{\DeclareRobustCommand{\VAN}[3]{##3}\VANthebibliography}

\usepackage{graphicx}	
\usepackage{amsmath}	

\newcommand{\z}[1]{\mbox {ZZ PsA}}

\def\p0{\phantom{0}}

\usepackage{xcolor}

\title[\z1\ and the instability mass ratio of contact binary stars]{ZZ Piscis Austrinus (\z1): A bright red nova progenitor and the instability mass ratio of contact binary stars}

\author[S. S. Wadhwa et al.]{Surjit S. Wadhwa,$^{1}$\thanks{E-mail: 19899347@student.westernsydney.edu.au}
Ain De Horta,$^{1}$
Miroslav D. Filipovi\'c,$^{1}$
N.~F.~H. Tothill,$^{1}$
\newauthor Bojan Arbutina,$^{2}$ Jelena Petrovi\'c$^{3}$ and Gojko Djura\v sevi\'c$^{3}$
\\
$^{1}$School of Science, Western Sydney University, Locked Bag 1797, Penrith, NSW 2751, Australia\\
$^{2}$Department of Astronomy, Faculty of Mathematics, University of Belgrade, Studentski trg 16, 11000 Belgrade, Serbia\\
$^{3}$Astronomical Observatory, Volgina 7, 11060 Belgrade, Serbia\\
}

\date{Accepted XXX. Received YYY; in original form ZZZ}

\pubyear{2020}

\begin{document}
\label{firstpage}
\pagerange{\pageref{firstpage}--\pageref{lastpage}}
\maketitle

\begin{abstract}
\z1\ is a neglected bright southern contact binary system with maximum V magnitude of 9.26. We present the first multi-band photometric analysis and find the system to be in deep contact (>95\%) with an extremely low mass ratio of 0.078. The primary has a mass of 1.213\,M$_{\sun}$ in keeping with its reported spectral class of F6. In order to determine if \z1\ is a merger candidate we outline the current status regarding the instability mass ratio and develop new relationship linking the mass of the primary to the instability mass ratio of the system and the degree of contact. We find that \z1\ along with two other examples from the literature to be merger candidates while an additional three require further observations to be confirmed as potential merger candidates.

\end{abstract}

\begin{keywords}
binaries: eclipsing -- stars: mass-loss -- techniques: photometric
\end{keywords}



\section{Introduction}
 \label{sec:intro}

Luminous red novae are rare transient events thought to be the result of the merger of contact binary system components into a single star. So far, there is only one confirmed observed event, that of V1309~Sco --- Nova~Sco~2008 \citep{2011A&A...528A.114T}. Unfortunately, that system was recognized as a contact binary only after the merger event, so targeted observations to fully elucidate the properties of the components and events leading up to the merger itself could not be performed. Estimates suggest that 1 in 500 stars in the galactic disk are contact binaries \citep{2006MNRAS.368.1319R}. Sky survey projects such as All Sky Automated Survey \citep[ASAS,][]{1997AcA....47..467P}, the Zwicky Transient Facility \citep[ZTF,][]{2019PASP..131a8002B}, Asteroid Terrestrial-impact Last Alert System \citep[ATLAS,][]{2018AJ....156..241H} and ASAS---SN \citep{2014ApJ...788...48S, 2018MNRAS.477.3145J}, among many others, are continually adding to our lists of contact binary systems. \citet{2014MNRAS.443.1319K} estimated the Galactic rate of stellar mergers such as V1309~Sco to be once every 10 years. Identification of candidates that are close to merging is critical to our understanding of these systems.

Most contact binary systems are of low mass, with stars of spectral class F to K \citep{2012AcA....62..153S}. Theoretical modelling suggests that several factors influence the rate at which any particular system moves towards merger. Among the most critical appear to be the relationship between orbital and spin angular momentum \citep{1995ApJ...444L..41R}, the degree of contact \citep{1995ApJ...438..887R} and angular momentum loss \citep{2012AcA....62..153S}.
 
The bright contact binary system ZZ~Piscis Austrinus (\z1) ($\alpha_{2000.0} = 21\ 50\ 35.17$, $\delta_{2000.0} = -27\ 48\ 35.48$) was recognized as a variable star in 1967 \citep{1967IBVS..195....1S}, ``rediscovered'' by \cite{1996IBVS.4322....1D}, and designated NSV~13890. Despite its brightness, the system has remained largely unobserved apart from sky surveys. ASAS confirmed the contact binary light curve with a period of 0.37388~days. A single time of minimum was reported by \citet{2004IBVS.5532....1O} based on the only section of the ASAS data of sufficient cadence to allow a rough estimation. Preliminary analysis of the ASAS photometry \citep{2006Ap&SS.301..195W} revealed a very low mass ratio system with a high degree of contact.

We present and analyse the first multi-band photometry of \z1 and determine its geometric and absolute parameters (Section~\ref{sec:LCA}). In Section~\ref{sec:4} we explore the problem of determining the orbital/spin angular momentum instability of contact binary systems and propose some new relationships between parameters determined by the light curve solution which shed new light on the subject of the instability mass ratio of contact binary systems. We apply the relationships to \z1\ and several other systems from the literature to define a very small number of systems that we suspect are potential red nova candidates. Finally, in Section~\ref{sec:summary} we summarise our results.

\section{Observations, Light Curve and Analysis}
 \label{sec:LCA}

\z1\ was observed over 5 nights during 2019 September--October with the Western Sydney University 0.6\,m telescope 
using standard Johnson $BVR$
filters. 
Images in all three bands were acquired during each observing session. Aperture photometry was performed on calibrated images using the American Association of Variable Star Observers (AAVSO) VPHOT engine. \mbox {TYC~6959--1122-1} and \mbox {TYC~6959--885--1} were used as comparison stars and \mbox {2MASS~21503231--2752141} was used as the check star.

\subsection{Period, Time of Minima and Light Curve Characteristics}

Our observations yielded a single time of primary minimum, determined to be \mbox {HJD~2458776.9669\ (0.0002)} using the method of \cite{1956BAN....12..327K}. A new period estimate of 0.37405 (0.00003) days was also determined using the PSearch V1.5 (Nelson 2006) software package available through the AAVSO. No meaningful period change (O-C) analysis was possible given the lack of previous observations as noted above. 

\begin{figure}
	\includegraphics[width=\columnwidth]{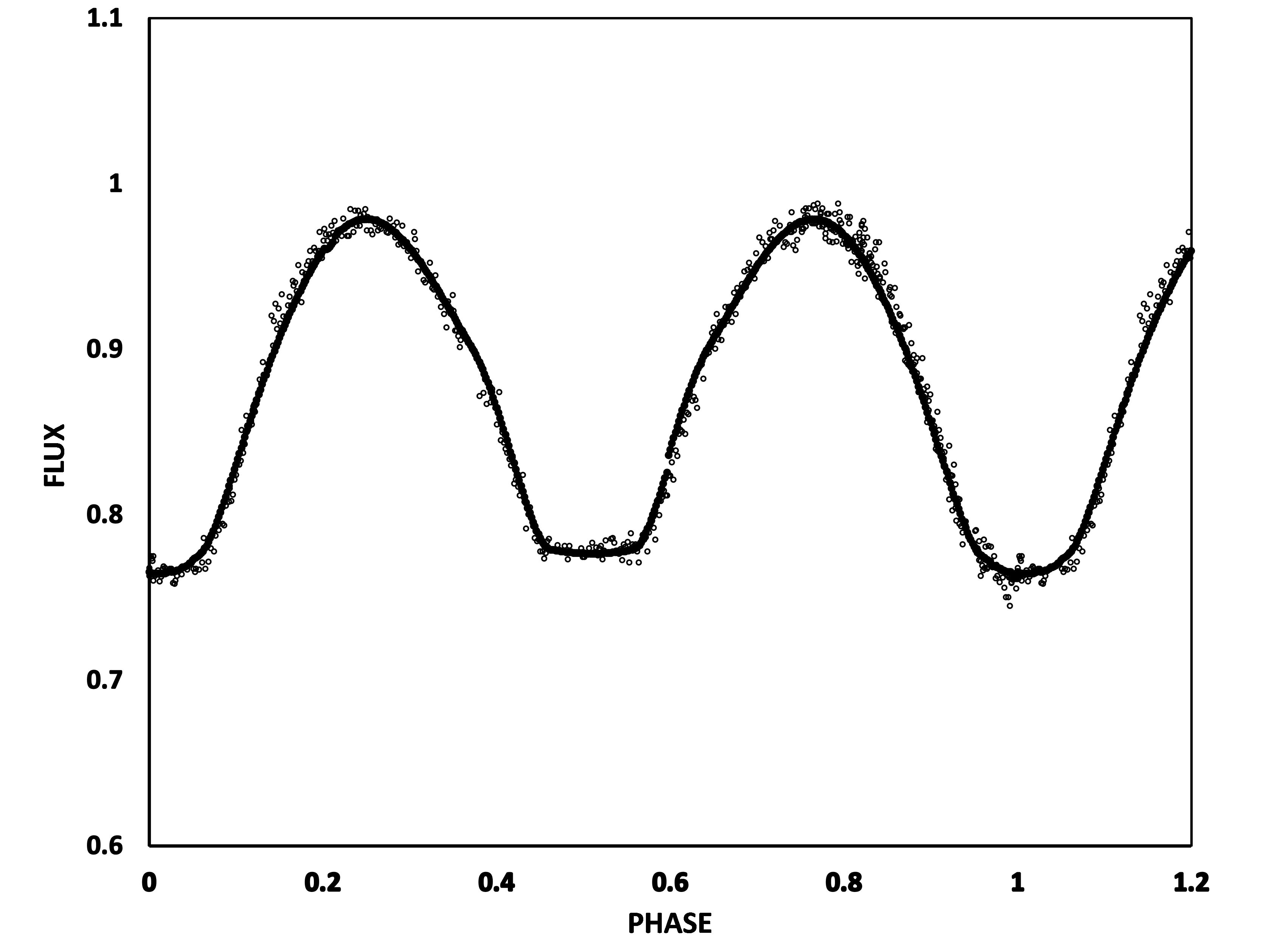}
    \caption{$B$ band light curve for \z1:
    Observed ($\circ$) and Simultaneous Fitted ($\bullet$) fluxes are shown.}
    \label{fig:tfig_1}
\end{figure}

\begin{figure}
	\includegraphics[width=\columnwidth]{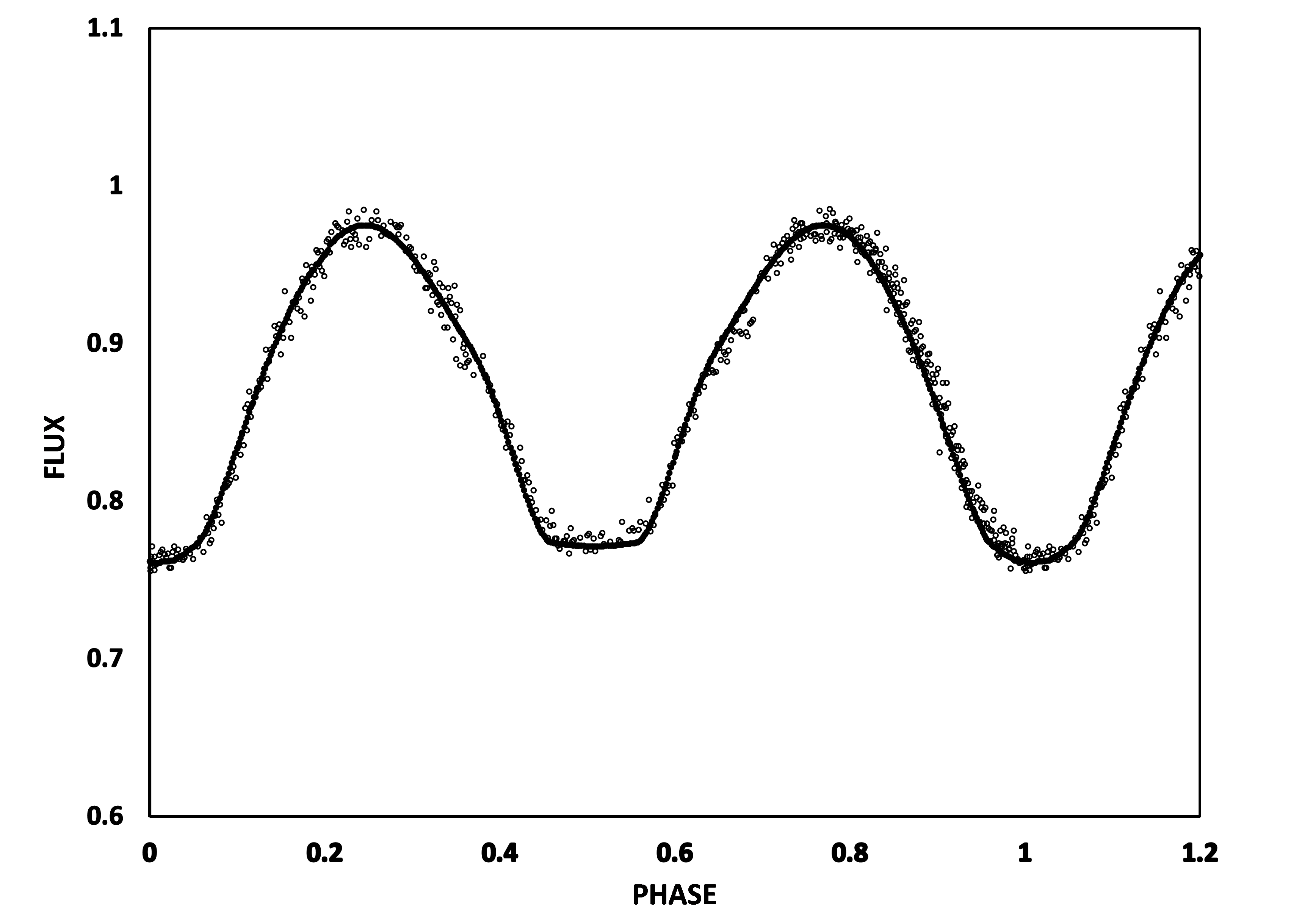}
    \caption{$V$ band light curve for \z1:
    Observed ($\circ$) and Simultaneous Fitted ($\bullet$) fluxes are shown.}
    \label{fig:tfig_2}
\end{figure}

\begin{figure}
	\includegraphics[width=\columnwidth]{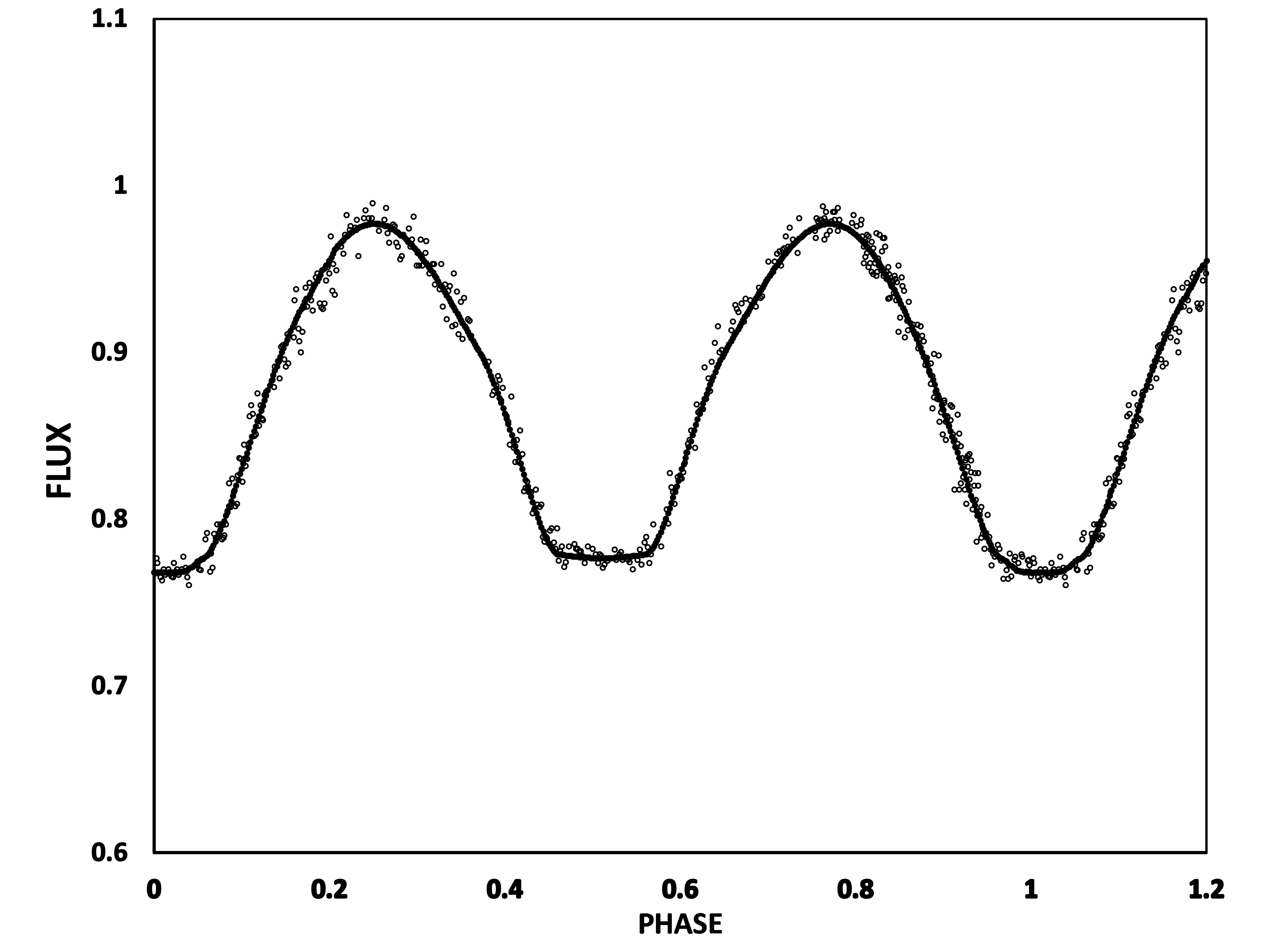}
    \caption{$R$ band light curve for \z1:
   Observed ($\circ$) and Simultaneous Fitted ($\bullet$) fluxes are shown.}
    \label{fig:tfig_3}
\end{figure}

The normalized and fitted light curves for all bands are shown in Figures~\ref{fig:tfig_1}, \ref{fig:tfig_2} and \ref{fig:tfig_3}. The light curves were folded according to the elements stated above. The maximum magnitudes at phase 0.25 and 0.75 in B were estimated as $9.83\pm 0.01$ and $9.84\pm 0.01$; in V~$9.26\pm 0.01$ and $9.26\pm 0.01$ and in R~$8.71\pm0.01$ and $8.72\pm 0.01$. Surprisingly, given the low mass ratio (see below) no significant variation in the maxima 
\citep[O'Connell effect, ][]{1968AJ.....73..708M}
was evident in any of the bands observed. The eclipses are of essentially equal depth (variation <~0.02\,mag) and last approximately 37~minutes. The $V$ band amplitude of the light curve is 0.29\,mag which agrees with the coarse ASAS survey amplitude of 0.27\,mag. \citet{2001AJ....122.1007R} developed a theoretical relationship between maximum eclipse amplitude, fill-out factor and the mass ratio of contact binary systems. Using values from Table 2 of that study we deduce that the system is likely to have a mass ratio below~0.1. 

\subsection{Photometric Analysis of the Light Curve}

The system shows complete eclipses and therefore it is suitable for photometric analysis with high confidence that the correct mass ratio can be determined \citep{2005Ap&SS.296..221T}. The light curve was analyzed using the 2009 version of the Wilson-Devinney (WD) code \citep{1971ApJ...166..605W}. As there is no appreciable O’Connell effect, only unspotted modelling was carried out. The SIMBAD database records the spectrum of the system as F6V \citep{2000A&AS..143....9W}. Based on calibrations from \citet{2000asqu.book.....C} the effective temperature of the primary component was fixed at 6514\,K. The near equal depth of the eclipses suggests similar temperatures of the components and a common convective envelope, as such we fixed the gravity darkening coefficients $g_1 = g_2 = 0.32$ \citep{1967ZA.....65...89L}, bolometric albedos $A_1 = A_2 = 0.5$, and applied simple reflection treatment \citep{1969PoAst..17..163R}. Limb darkening co-coefficients for each passband were obtained from \citet{1993AJ....106.2096V} and logarithmic law for the coefficients was implemented as per \citet{2015IBVS.6134....1N}. 
 
No spectroscopic mass ratio is available, only the visual classification (F6V) by \citet{1982mcts.book.....H} appears in the literature. The complete eclipses however mean that the photometric mass ratio can be more reliably discovered and is likely to be close to the actual mass ratio \citep{2005Ap&SS.296..221T}. 
\citet{1993PASP..105.1433R} showed that the brightness variations of the contact binary stars are essentially completely dominated by geometrical elements, with reflection and gravity/limb darkening playing a minimal role. He showed that the light curve depends practically on only three parameters; the mass ratio $(q)$, inclination $(i)$ and the degree of contact or fill-out $(f)$. Unfortunately, there is a high degree of correlation between these parameters. 
The correlation is most marked between mass ratio and inclination, and between mass ratio and degree of contact. Reliable photometric solutions cannot be obtained when either of these two combinations are treated as free. \citet{1982A&A...107..197R} described the ``$q$'' search or ``grid'' method whereby solutions are obtained for various fixed values of $q$, with the potential (fill-out) and inclination free and then compared to the observed light curve to find the best solution. \citet{2006Ap&SS.301..195W} obtained a preliminary mass ratio of 0.08 based on the analysis of ASAS data. 

\begin{figure}
	\includegraphics[width=\columnwidth]{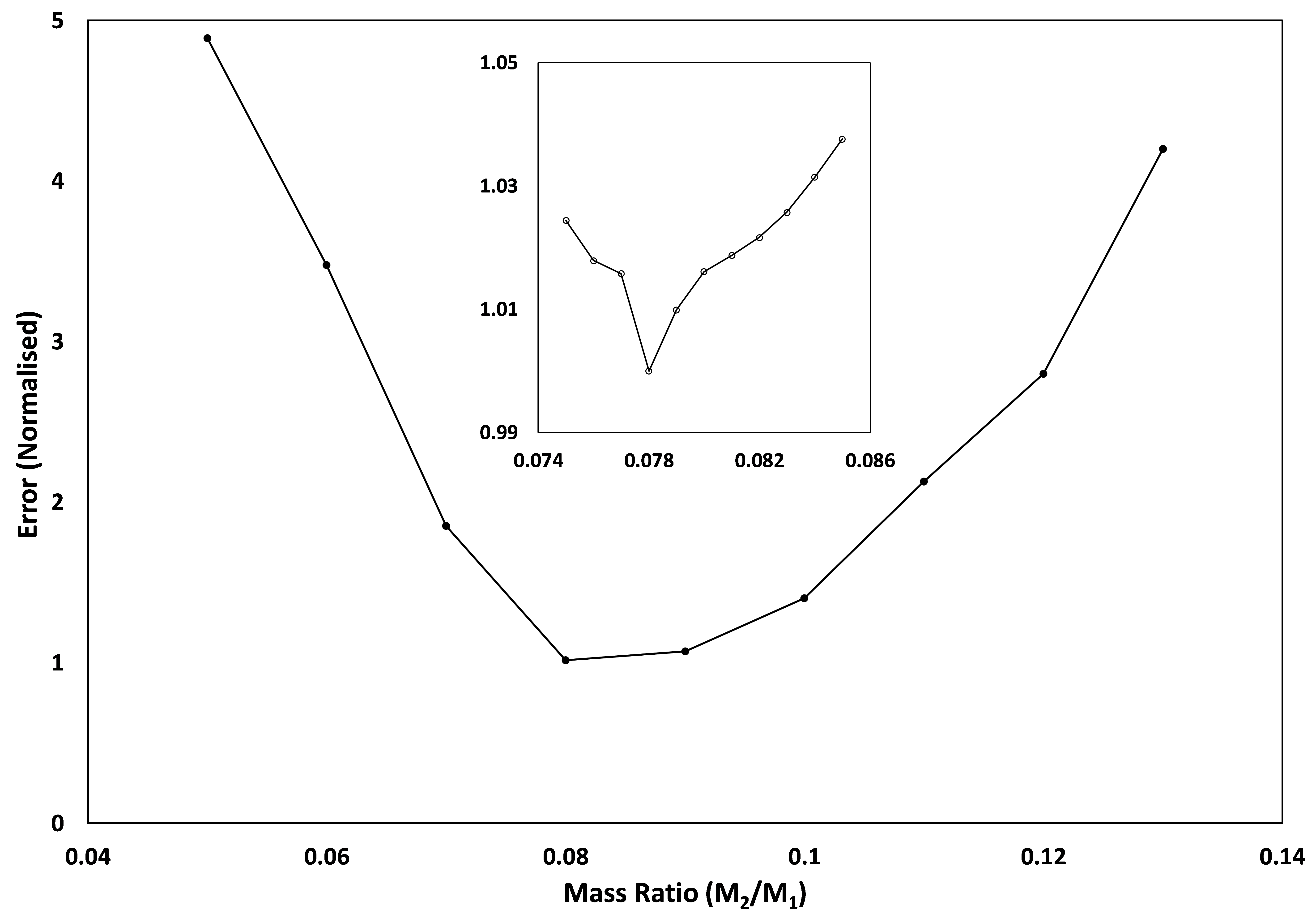}
    \caption{Coarse and fine (inset) mass ratio (q) search grid for \z1.}
    \label{fig:tfig_4}
\end{figure}

\begin{figure}
	\includegraphics[width=\columnwidth]{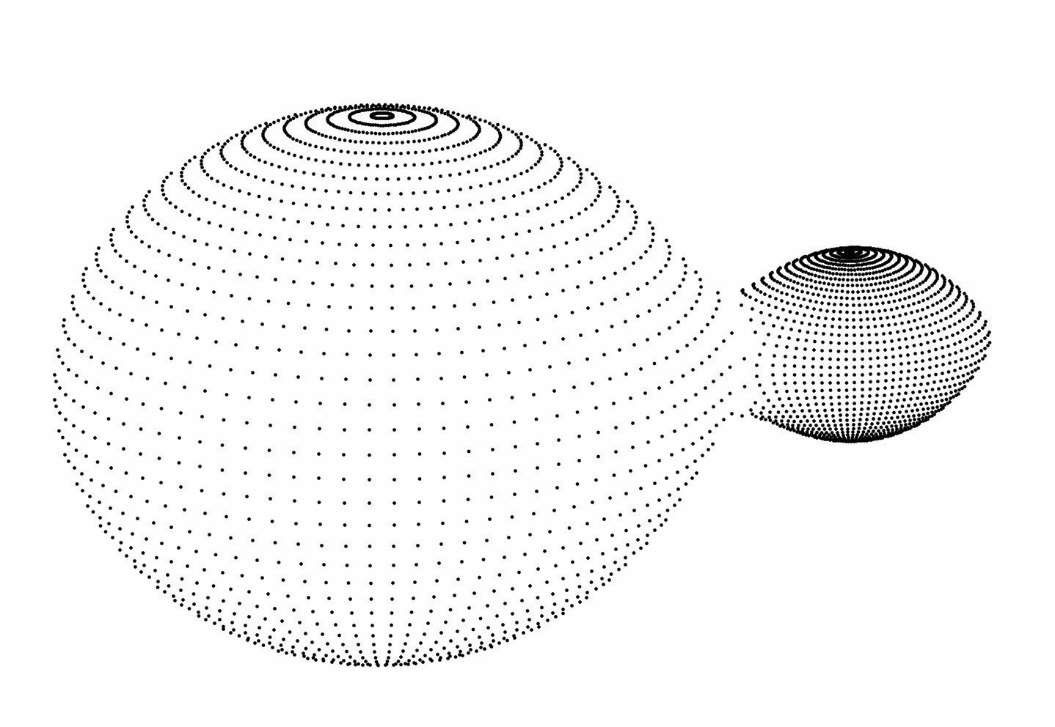}
    \caption{3D Representation of \z1 at phase 0.35. The broad neck between the components is an indicator of the high fill-out.}
    \label{fig:tfig_5}
\end{figure}

Using the WDWint package V5.6 (Nelson 2009) we analyzed the light curves of all three passbands simultaneously over the mass ratio range 0.05 to 0.13 at intervals of 0.01. Preliminary solution was found at $q = 0.08$. To further refine the mass ratio a finer search at intervals of 0.001 for the mass ratio was made over the interval $0.075 < q < 0.085$. Throughout the analysis the inclination, potential (fill-out), temperature of the secondary and relative value of $L_1$ were adjusted for each fixed value of the mass ratio. Iterations were carried out until the suggested variation was less than the reported standard deviation. The normalised residual error (as reported by the analysis software) between the observed and modelled light curve was plotted against the mass ratio. The mass ratio with the smallest residual error, i.e. the best fit between observed and modelled light curve, was selected as the true mass ratio of the system. The ``$q$'' search grid is shown in Figure~\ref{fig:tfig_4} and the critical parameters from the WD solution are presented in Table~\ref{tab:table_1}. A 3D representation of the system is shown in Figure~\ref{fig:tfig_5}.

\subsection{Review of the Light Curve Solution and Determination of Absolute Parameters}

 Our light curve solution, with 
 an estimated fill-out of >95\%, represents one of the highest recorded degrees of contact for a binary system and suggests probable near-merger. In addition to good physical contact, there is solid thermal contact between the systems. The secondary, although smaller than the primary, is hotter by less than 200\,K and as such the system is similar to other low mass ratio systems with hotter secondaries such as V857~Her \citep{2005MNRAS.356..765Q}, \mbox{ASAS~J083241+2332.4} \citep{2016AJ....151...69S}, \mbox{V1187~Her} \citep{2017JAVSO..45...11W}, {LO~And} \citep{2015IBVS.6134....1N} and {TY~Pup} \citep{ 2018AJ....156..199S}.

\begin{table}
	\centering
	\caption{Key parameters from the WD solution for \z1.}
	\label{tab:table_1}
	\begin{tabular}{cc} 
		\hline
		$T_1$ (K) (Fixed) & 6514 \\
		$T_2$ (K) & $6703 \pm 11$ \\
		$q({M}_2$/${M}_1)$ & $0.078 \pm 0.002$ \\
		$\text{Inclination }$($^{\circ}$)& $75.25\pm 0.34$ \\
				$\Omega_\text{1}=\Omega_\text{2}$ & $1.839 \pm 0.001$ \\
		Fill-out\% & $97 \pm 2$ \\
		\hline
	\end{tabular}
\end{table}

Apart from the mass ratio, determination of the absolute magnitude $\text{M}_\text{v}$ of the primary is essential to accurately determine the absolute parameters of the system. Based on previously published colour-period and colour-period-$\text{M}_\text{v}$ relationships, \citet {2008MNRAS.390.1577G} derived the relationship $\text{M}_\text{v} = -8.4\log P + 0.31$ (where $P$ is the period in days), giving $\text{M}_\text{v} = 3.90$ for the \z1\ primary. However, the distance to \z1\ is known from parallax to be $135.48\pm 0.99$~pc \citep{2018yCat.1345....0G}. During the eclipse at phase 0.5 (lasting about 37~minutes), only the light from the primary contributes to the light curve at mid-eclipse. The apparent visual magnitude at mid-eclipse was estimated as $9.53\pm 0.01$, giving a distance-based absolute magnitude of $3.97\pm 0.02$ for the primary, after applying the extinction correction of $E(B-V) = 0.034$ \citep {1998ApJ...500..525S}, as described by \citet{2019JAVSO..47..138W}. The distance-based results are used in all subsequent calculations. 

From the primary $\text{M}_\text{v}$, we can find the primary luminosity. At maximum brightness (phase 0.25 or 0.75), the system has $V= 9.26\pm 0.01$ which yields system absolute magnitude M$_V = 3.70\pm 0.02$, and hence the system luminosity, secondary luminosity and secondary absolute magnitude. \citet{2001Ap&SS.275..337R} found that the primary components of contact binary systems follow in general the zero-age main sequence (ZAMS) mass-luminosity relation. Using this relation \citep{1991Ap&SS.181..313D} for ZAMS stars with mass $>$ 0.7$M_{\sun}$, we can derive the mass of the primary; from the mass ratio we then estimate the mass of the secondary. The estimated $M_\text{1}$ of 1.213$M_{\sun}$ is consistent with the F6 spectral classification. The separation of the components $A$ is then estimated from Kepler's third law, expressed as $(A/R_{\sun})^3=74.5(P/\mathrm{days})^2((M_1/M_{\sun})+(M_2/M_{\sun}))$. 

As the two components are enclosed in a common envelope, their relative radii are not independent, but dependent on the Roche geometry and mass ratio. The WD program solution provides the fractional radii of the primary and secondary in three orientations. From the geometric mean we estimate the fractional radius of the primary $r_1=0.5958$, and that of the secondary $r_2=0.2341$. From ${R}_1/{R}_{\sun} = r_1 A$ and ${R}_2/{R}_{\sun} = r_2 A$ \citep{2005JKAS...38...43A},  we derive the primary and secondary radii. 
The relative radius and luminosity of the secondary are considerably higher than those predicted for stars either at ZAMS or terminal-age main sequence (TAMS) stages of evolution. This is a common finding among contact binary systems \citep{2012IAUS..282..456S}. The absolute parameters for \z1\ are given in Table~\ref{tab:table_2}.

\begin{table}
	\centering
	\caption{Absolute parameters for \z1:}
	\label{tab:table_2}
	\begin{tabular}{cc} 
		\hline
		${M}_1/{M}_{\sun}$ & $1.213\pm0.008$ \\
		${M}_2/{M}_{\sun}$ & $0.095\pm0.001$ \\
		${L}_1/{L}_{\sun}$ & $2.20\pm 0.04$ \\
		${L}_2/{L}_{\sun}$ & $0.63\pm 0.04$ \\
		${R}_1/{R}_{\sun}$ & $1.422\pm 0.004$\\
		${R}_2/{R}_{\sun}$ & $0.559\pm 0.004$ \\
		$\text{A}~({R}_{\sun})$ & $2.387\pm 0.008$ \\
		$\text{M}_{\text{V}1}$ & $3.97\pm 0.02$ \\ 
		$\text{M}_{\text{V}2}$ & $5.34\pm 0.02$ \\
			\hline
	\end{tabular}
\end{table}

\section{Instability mass ratio of Contact Binary Systems}
\label{sec:4}

Contact binary systems are thought to merge into a single, rapidly spinning object when the total angular momentum ($J_{T}$) of the system is at a critical ($dJ_T=0$) value such that tidal instability ensues \citep{1879Obs.....3...79D}. $J_{T}$ is the sum of the orbital angular momentum ($J_{O}$) of the system and the combined spin angular momentum ($J_{S}$) of the components. The instability occurs $\approx$ when $J_{O}\leq3J_{S}$ \citep{1980A&A....92..167H}. For contact binaries, this equates to a critical mass ratio of 0.071--0.077 \citep{1995ApJ...444L..41R}, depending on the dimensionless gyration radius $k$ of the component stars and the degree of contact. Over the years, $k$ appears to have been chosen empirically to match the observations. \citet{1995ApJ...444L..41R} chose $k=0.245$, such that the lowest mass ratio system known at the time, AW~UMa with $q=0.075$ \citep{1992AJ....104.1968R} would be stable, and did not take into account the spin momentum of the secondary when arriving at the global critical mass ratio. As more systems with mass ratios below the estimated critical value were discovered, researchers tried to modify the calculations either to include the spin momentum of the secondary \citep{2006MNRAS.369.2001L} or to use different values of $k$ \citep{2010MNRAS.405.2485J}. 

With differing values of $k$ being employed, there is no established relationship to confidently say whether a contact binary system is unstable regardless of its measured mass ratio.  \citet{1988AJ.....95.1895R} and \citet{2019A&A...628A..29C} calculated $k$ for stars of different masses and radii (non-rotating), showing that there is considerable variation in $k$ with changing stellar mass. In both cases, stars with mass 
$>0.8 M_{\sun}$ were modelled and no rotational or tidal effects were included. \citet{2009A&A...494..209L} determined gyration radii for ZAMS stars, including effects of rotation and tidal distortion due to binary interaction. We use these results to derive  more robust and complete criteria for the critical instability mass ratio of contact binary systems with different masses of the primary component.

\subsection{The Instability Mass Ratio ($q_{inst}$)}
Our derivation of the instability mass ratio is similar to that of \citet{2007MNRAS.377.1635A}, expanded to more clearly show 
the contribution of the secondary component. We adopt the stability criterion $dJ_{T} = 0$, assume that the gyration radii of the primary and secondary are not equal, and further assume that, as both stars are within a common envelope, their radii are not independent, but the radius of the secondary is dependent on the separation, mass ratio and the radius of the primary.

The orbital angular momentum of a binary can be written as
\begin{equation}
J_{\mathrm{O}} =  \mu A^2 \omega = \frac{q\ \sqrt{G M^3
A}}{(1+q)^2},
\end{equation}
where $M_1$ and $M_2$ are masses of the primary and secondary component,
respectively, total mass $M=M_1+M_2$, $\mu = M_1 M_2 /M$ and $q = M_2/M_1 < 1$.
$\omega$ is the orbital angular velocity.
With synchronization assumed, the spin angular momentum of a binary
\begin{equation}
J_{\mathrm{S}} =  k_1^2 M_1 R_1^2 \omega + k_2^2 M_2 R_2^2
\omega,
\end{equation}
where $R_1$ and $R_2$ are taken to be the volume radii.
The fill-out factor is defined as
\begin{equation}
\label{eq:define-f}
f =  \frac{\Omega - \Omega _{\mathrm{IL}}}{\Omega _{\mathrm{OL}} - \Omega
_{\mathrm{IL}}} \approx \frac{R - R _{\mathrm{IL}}}{R
_{\mathrm{OL}} - R _{\mathrm{IL}}},
\end{equation}
where $\Omega$ is the dimensionless surface equipotential.\
We have adopted linear dependence of $\Omega$ on the volume
radius \citep{2007MNRAS.377.1635A}. Volume
radii for the inner Roche lobe are \citep{1983ApJ...268..368E}:
\begin{large}
\begin{equation}
  \frac{R_{\mathrm{IL}i}}{A} = \Bigg\{ \begin{array}{ll}
 { \frac{0.49q^{-2/3}}{0.6q^{-2/3} + \ln (1+ q^{-1/3})}, } &  i=1 \\\\
 { \frac{0.49q^{2/3}}{0.6q^{2/3} + \ln (1+ q^{1/3})}, }  &  i=2,
 \end{array}
\end{equation}
\end{large}
and for the outer Roche lobe \citep{2005ApJ...629.1055Y}:
\begin{large}
\begin{equation}
  \frac{R_{\mathrm{OL}i}}{A} = \Bigg\{ \begin{array}{ll}
 { \frac{0.49q^{-2/3}+0.15}{0.6q^{-2/3} + \ln (1+ q^{-1/3})}, } & i=1 \\\\
  { \frac{0.49q^{2/3}+ 0.27q -0.12q^{4/3}}{0.6q^{2/3} + \ln (1+ q^{1/3})}.
 } & i=2,
\end{array}
\end{equation}
\end{large}
\begin{flushleft}
where $i=1$ and $2$ refer to the primary and secondary star.
\end{flushleft}
Since the surfaces of components in the contact system are at the same
potential (the same $f$), from Eq.~\ref{eq:define-f}:
\begin{equation}
R_2 = P(q)A + Q(q)R_1,
\end{equation}
where
\begin{footnotesize}
\begin{equation}
Q(q) = \frac{R_{\mathrm{OL}2} - R_{\mathrm{IL}2}}{R_{\mathrm{OL}1}
- R_{\mathrm{IL}1}}\\=\frac{(0.27q -0.12q^{4/3})({0.6q^{-2/3} + \ln (1+ q^{-1/3})})}{0.15 (0.6q^{2/3} + \ln (1+ q^{1/3}))},
\end{equation}
\end{footnotesize}
and
\begin{footnotesize}

\begin{equation}
P(q) = \frac{R_{\mathrm{IL}2}}{A} - Q(q)
\frac{R_{\mathrm{IL}1}}{A} = \frac{0.49q^{2/3}-3.26667q^{-2/3}(0.27q -0.12q^{4/3})}{0.6q^{2/3} + \ln (1+ q^{1/3})} .
\end{equation}
\end{footnotesize}
\begin{flushleft}
The total angular momentum of a contact binary system
$J_{\mathrm{T}} = J_{\mathrm{O}} + J_{\mathrm{S}}$ can then be expressed as
\end{flushleft}
\begin{equation}
\begin{array}{@{\extracolsep{-3mm}} lll @{}}
{  J_{\mathrm{tot}} } &=&  \frac{q\ \sqrt{G M^3 R_1}}{(1+q)^2}
\Big(\frac{A}{R_1}\Big)^{1/2}
                  \bigg[ 1 + \ \\
                   &+&
               \frac{k_1^2 (1+q)}{q}
               \Big( (1+ q \frac{k_2^2}{k_1^2}{Q}^2) \Big( \frac{R_1}{A}  \Big)^2  + 2q \frac{k_2^2}{k_1^2}{P}{Q} \Big( \frac{R_1}{A}  \Big) + q \frac{k_2^2}{k_1^2}{P}^2 \Big)
               \bigg] ,
\end{array}
\end{equation}
\
\begin{flushleft}
From the condition $\frac {\mathrm{d}
J_{\mathrm{T}} }{\mathrm{d} A}=0$, the critical separation ($A_{inst}$) is given by
\end{flushleft}
\begin{small}

\begin{equation}
\label{eq:a-inst}
 \frac{A_{\mathrm{\scriptscriptstyle inst}}}{R_1} = \frac{q\frac{k_2^2}{k_1^2}{P}{Q} + \sqrt{(q\frac{k_2^2}{k_1^2}{P}{Q})^2 + 3 (1+q\frac{k_2^2}{k_1^2}{Q}^2) (q \frac{k_2^2}{k_1^2}{P}^2 + \frac{q}{(1+q)k_1^2})   }}{q \frac{k_2^2}{k_1^2}{P}^2 + \frac{q}{(1+q)k_1^2}}.
\end{equation}
\end{small}
\begin{flushleft}
If the angular momentum of the secondary has been neglected ($k_2 = 0$), the above condition for critical separation reduces to \citep{ 1995ApJ...444L..41R}\end{flushleft}
\begin{equation}
\label{eq:ainst-for-k0}
 \frac{A_{\mathrm{\scriptscriptstyle inst}}}{R_1} = k_1 \sqrt{\frac{3(1+q)}{q}}.
\end{equation}
\begin{flushleft}
By combining Eq.~\ref{eq:define-f} for the primary\end{flushleft}
\begin{equation}
\label{eq:R1/A}
 \frac{R_1}{A} = f \Big(\frac{R_{\mathrm{OL}1}}{A} - \frac{R_{\mathrm{IL}1}}{A}\Big) + \frac{R_{\mathrm{IL}1}}{A} = \frac{0.49q^{-2/3}+0.15 f}{0.6q^{-2/3} + \ln (1+ q^{-1/3})} 
\end{equation}
and Eq.~\ref{eq:a-inst}, for a fixed fill-out factor ($0\leq f\leq1$) one can also find the instability mass ratio given in the implicit form by an algebraic equation that needs to be solved numerically for $q \equiv q_{\mathrm{\scriptscriptstyle inst}}$:
\begin{equation}
\begin{split}
\begin{array}{@{\extracolsep{-3mm}} lll @{}}
\frac{q\frac{k_2^2}{k_1^2}{P}{Q} + \sqrt{(q\frac{k_2^2}{k_1^2}{P}{Q})^2 + 3 (1+q\frac{k_2^2}{k_1^2}{Q}^2) (q \frac{k_2^2}{k_1^2}{P}^2 + \frac{q}{(1+q)k_1^2})   }} {q \frac{k_2^2}{k_1^2}{P}^2 + \frac{q}{(1+q)k_1^2}} &
 - \\ \\ \frac{0.6q^{-2/3} + \ln (1+ q^{-1/3})}{0.49q^{-2/3}+0.15 f}  = 0.
 \end{array}
 \end{split}
\end{equation}

Given the low mass ratio of contact binary systems approaching merger, the secondary is a very low mass star, for which the totally convective $n=1.5$ polytrope would be appropriate, with $k^2\approx 0.205$. The value is almost identical to the value obtained for all ZAMS stars below 0.4\,M$_{\sun}$ modelled by \citet{2009A&A...494..209L}. Our relationships are derived by adopting this fixed value of $k$ for the secondary. We use the values of $k$ for 0.5\,M$_{\sun}$ to 1.6\,M$_{\sun}$ from \citet{2009A&A...494..209L} to derive the instability mass ratio ($q_{inst}$), for a range of fill-out fractions, of contact binary systems with primary component of spectral classes from M0 to F0. The results are summarised in Table 3. The fill-out fraction $f$ has a significant effect on $q_{inst}$ for systems with primary components below one solar mass. On the other hand, there is minimal difference in $q_{inst}$ between $f=1$ and $f=0$ for systems with primaries in the upper F spectral class. Figure 6 shows the relationship between primary mass and $q_{inst}$ at $f=1$ and at $f=0$. Both can be fitted with quadratic relationships (correlation coefficients >0.99):
\begin{equation}
\label{eq:qinst-f1}
    q_{inst}=0.1269M_{1}^2-0.4496M_{1}+0.4403 (f=1)
\end{equation} 
and
\begin{equation}
\label{eq:qinst-f0}
  q_{inst}=0.0772M_{1}^2-0.3003M_{1}+0.3237 (f=0).  
\end{equation}

\begin{table}
\label{tab:mass-ratios}
	\centering
	\caption{Instability mass ratio for systems with primary stars of $0.5M_{\sun}<M<1.6_{\sun}$ and various fill-out fractions. $q_{inst} $ as $M_2/M_1$ }
	\label{tab:table_3}
		\begin{tabular}{ccccccc} 
		\hline
		Mass & $q_{inst}$ & $q_{inst}$ & $q_{inst}$ & $q_{inst}$ & $q_{inst}$ & $q_{inst}$\\
		($M_{\sun}$) & $f1.0$ & $f0.9$ & $f0.75$ & $f0.5$ & $f0.25$ & $f0$\\
		\hline
		0.5& 0.253&	0.246&	0.236&	0.220&	0.207&	0.196\\
        0.6	&0.213&	0.208&	0.200&	0.189&	0.178&	0.169\\
        0.7	&0.182&	0.178&	0.172&	0.163&	0.155&	0.148\\
        0.8	&0.160&	0.156&	0.152&	0.145&	0.138&	0.132\\
        0.9 &0.136&	0.134&	0.130&	0.125&	0.120&	0.115\\
        1.0	&0.126&	0.124&	0.121&	0.116&	0.111&	0.107\\
        1.2	&0.087&	0.086&	0.084&	0.081&	0.079&	0.076\\
        1.4	&0.054&	0.053&	0.052&	0.051&	0.050&	0.049\\
        1.6	&0.047&	0.046&	0.046&	0.045&	0.044&	0.043\\
		\hline
	\end{tabular}
\end{table}

    \begin{figure}
    \label{fig:qinst}
	\includegraphics[width=\columnwidth]{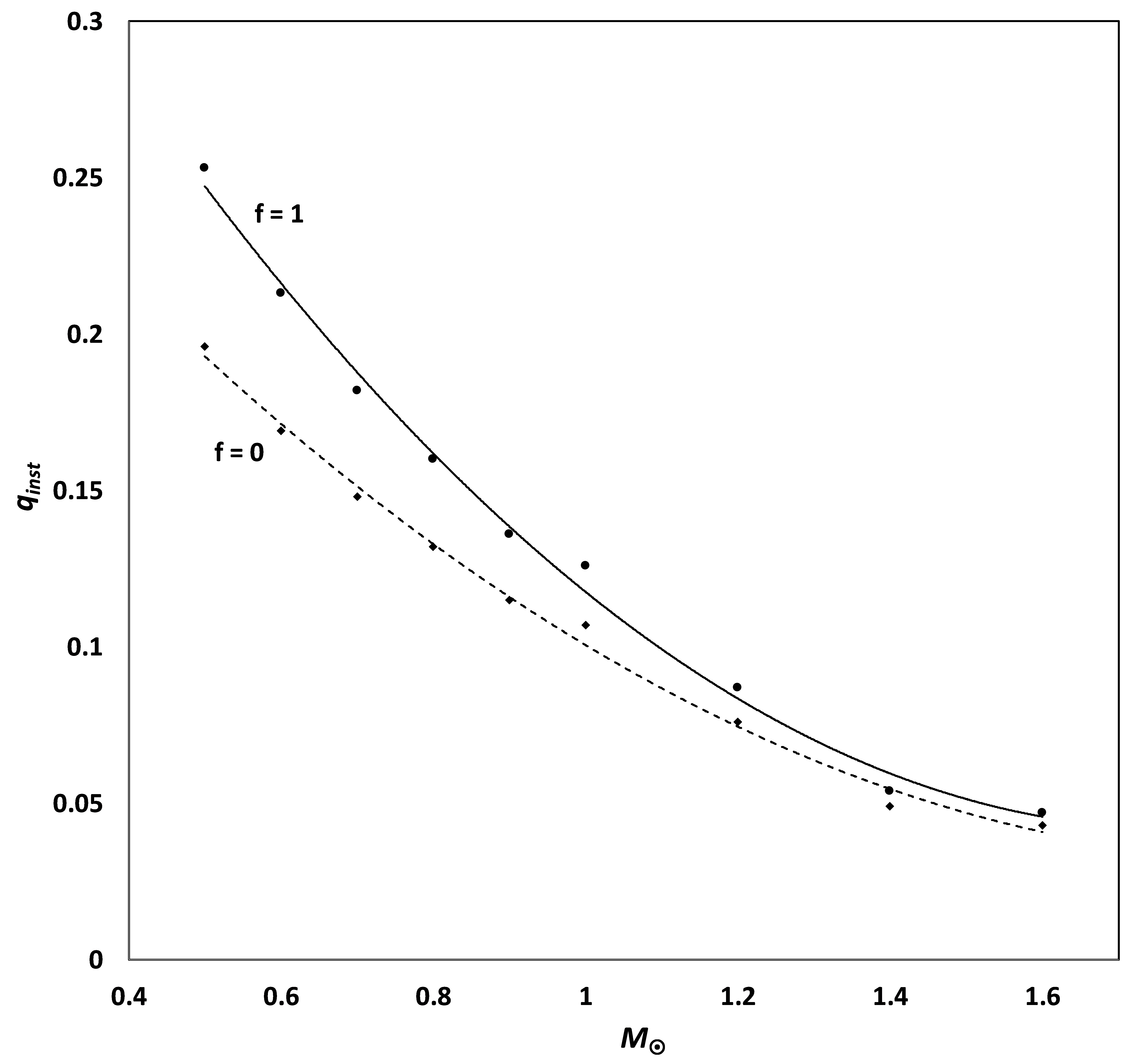}
    \caption{Instability mass ratios for  0.5\,M$_{\sun}$ to 1.6\,M$_{\sun}$ stars at $f=1$ and $f=0$}
    \label{fig:tfig_6}
\end{figure}

Table 3 shows a linear relationship between 
$f$ and $q_{inst}$ for stars of different mass. One can therefore easily obtain the instability mass ratio for any given fill-out, after obtaining the values for $f=1$ and $f=0$ using the above quadratic relationships.

 \subsection{Instability Parameters of ZZ PsA and Other Merger Candidates}

Using Eqs.~\ref{eq:qinst-f1} and \ref{eq:qinst-f0}, we estimate the instability mass ratio of \z1\ to be 0.0813. The photometric mass ratio is marginally below this level, suggesting that the system has entered an unstable period. From Eq~\ref{eq:a-inst} the theoretical instability separation can be expressed as a function of the mass ratio, primary radius and the gyration radii. Using the photometric mass ratio for \z1\ and Eq~\ref{eq:R1/A}, we determine the theoretical fractional radius of the primary as 0.630 which at the current separation would yield a radius for the primary as 1.50\,$R_{\sun}$. Using this estimate of the radius of the primary and 0.2432 as the value of $k$ from \citet{2009A&A...494..209L} for a 1.2\,$M_{\sun}$ star, from equation 10 we derive the theoretical instability separation as 2.487\,$R_{\sun}$. Combining the theoretical instability separation with our component mass estimates, we would expect that the system reach its instability separation at a period of 0.397 days. \z1 is well within the instability parameters on a number of fronts and represents a very bright potential red nova progenitor worthy of further study and regular monitoring. 

Since the observed merger event of V1309~Sco, there has been considerable interest in the analysis of contact binary light curves. We examined the list of 46 deep low mass ratio contact systems from \citet{2015AJ....150...69Y} and searched the literature for more recent examples of potential merger candidates. We used the reported values of the light curve solutions and mass of the primary to exclude all systems where the reported value of $q$ was more than 5\% above $q_{inst}$ as calculated using the relationships above. For the  remaining examples the instability separation and estimated period at onset of instability were compared to the current reported values. From an initial sample of more than 60 systems we were left with two other potential merger candidates and three others that were difficult to classify. We discuss these briefly.

\subsubsection{V857~Her}

V857~Her was observed by \citet{2005AJ....130.1206Q} and found to be a W-Type system with  mass ratio 0.0653, a fill-out of 83.8\%, and the secondary hotter than the primary (very similar to \z1). \citet{2015AJ....150...69Y} reported the mass of the primary as $1.3M_{\sun}$.  The instability mass ratio for such a system would be 0.069. Based on the reported period of 0.3822 days we estimate the current separation as 2.473$R_{\sun}$. The fractional radius of the primary is estimated as 0.6369 and from the current separation the radius of the primary is estimated as 1.575$R_{\sun}$. Although \citet{2009A&A...494..209L} did not specifically report the gyration radius for a 1.3$M_{\sun}$ star, review of the reported values suggest a linear decline from 0.6$M_{\sun}$ to 1.4$M_{\sun}$, so we estimate the gyration radius of the primary of V857~Her as 0.216. Using Eq 10, we determine the separation distance at instability to be 2.476$R_{\sun}$ and the period at onset of instability to be 0.3829 days. Just like \z1 V857~Her appears to be unstable, and a potential red nova progenitor.

\subsubsection{V1187~Her}

Preliminary analysis of amateur red band photometry by \citet{2017JAVSO..45...11W} suggested a W-Type star with mass ratio 0.059 and only marginal fill-out of 20\%. Multi-band photometry analysis \citep{2019PASP..131e4203C}, however, yielded a system with an extremely low mass ratio of 0.044 and fill-out exceeding 80\%. Although no component masses were givenn, the spectrum is reported to be F8.5, corresponding to a primary mass of  $1.1\,M_{\sun}$ \citep{2000asqu.book.....C}. Hence we derive $q_{inst}= 0.097$ and, from  other parameters in the light curve solution, the separation $A=2.02\,R_{\sun}$. The system is so far below the estimated critical mass ratio that it should have already merged. Photometric mass ratios are not as reliable as spectroscopic observations particularly in systems with low inclination, as stressed by \citet{2008MNRAS.388.1831R} in their analysis of GSC~1387:0475 and confirmed by \citet{2010NewA...15..155Y} who derived a significantly different photometric solution for the same system. V-1187 Her has a very low inclination (66$^{\circ}$).
This system is very likely to be unstable, but further spectroscopic studies are probably required to fully elucidate the parameters.

\subsubsection{EM~Psc, FG~Hya, and V870~Ara}

EM~Psc was observed by \citet{2005Ap&SS.300..337Y} who found it be a W-Type contact binary with a mass ratio of 0.1 and fill-out $>$90\%. \citet{2008AJ....136.1940Q} observed the system and found it to be an A-Type with a mass ratio of 0.149 and fill-out $>$90\%. Neither study estimated the mass of the primary. \citet{2015AJ....150...69Y}, in their statistical analysis of 46 low mass ratio systems, indicate the mass of the primary as 0.95$M_{\sun}$. At the reported mass of the primary the instability mass ratio is estimated as 0.126. If we take the mass ratio as 0.149 than the system would not be considered unstable. If, however we were to adopt the mass ratio of 0.1 from \citet{2005Ap&SS.300..337Y} than the system would be well below the critical mass ratio and a potential merger candidate. The system clearly needs further study. 

FG~Hya was observed by \citet{2005MNRAS.356..765Q} who found mass ratio 0.11 with $>$85\% contact. The only available value for the mass of the primary $1.41M_{\sun}$ is from \citet{1999AJ....118..515L}, who stressed that the value was likely to be erroneous. If 
this primary mass is adopted, the estimated critical mass ratio is 0.059, and the system would not be classified as unstable. The spectral classification of the system however is estimated as G0 \citep{2005MNRAS.356..765Q}, and if we assume that the primary stars of contact binaries follow the ZAMS, this would yield a primary mass of $1.05M{\sun}$ and a  critical mass ratio 0f 0.106, placing the system on the border of instability. 

Lastly, a combined photometric and spectroscopic solution for V870~Ara \citep{2007A&A...465..943S} gives mass ratio 0.082 and a primary mass of $1.50M_{\sun}$. On this basis, the system is stable. The spectral class, however, was determined to be F8 --- far later than the mass of the primary. The system would be unstable if the mass of the primary based on spectral class was adopted.

\subsubsection{Some Recent Reports}

Since the early drafts of this paper, a recent study summarising light curve solutions from survey photometry of the northern Catalina Sky Survey \citep{2020ApJS..247...50S} has been published. Physical parameters of over 2300 late type contact binary systems were derived from the analysis of the survey photometry. Our preliminary analysis suggests that $\sim 25$ candidates may fit the mass ratio criteria for instability. As this analysis was carried out on survey photometry, the identification of such systems opens the window for dedicated multi-band photometry of targeted systems that may prove to be merger candidates.

\section{Summary and Conclusions}
\label{sec:summary}

We present the first multi-band photometric analysis for the bright southern contact binary \z1. We find that the system has an extremely low mass ratio of 0.078 and high degree of contact, in excess of 95\%.  With the distance well-established by parallax, we were able to accurately determine the absolute parameters of the components, and of the system as a whole. The primary exhibits properties similar to other ZAMS stars of the same spectral type, while the secondary, as is typical in contact binary systems, has a larger radius and luminosity than would be expected. 

In order to determine if \z1\ is a potential merger candidate, we explore the concept of the instability mass ratio of contact binary systems and find it currently sub-optimal because of the empirical values used for certain parameters. Based on several established models of the internal stellar geometry, we propose a new concept of an instability mass ratio (as opposed to the minimum mass ratio), which is dependent on the mass of the primary and modelled with variable values of the gyration radius rather than the empirically fixed values previously employed. Our relationships suggest that within the spectral range F0 to M0 (and most contact binary systems fall within this range) earlier spectral class systems are likely to become unstable at extremely low mass ratios while the later types may become unstable at mass ratios near 0.25. We show that ZZ PsA is likely a highly unstable contact binary system and hence a potential merger (red nova) candidate. We identify two other potential merger candidates in the literature and further three which require ongoing study to confirm instability.

\section*{Acknowledgements}

JP and GD acknowledge the financial support of the Ministry of Education, Science and Technological Development of the Republic of Serbia through contract No 451-03-68/2020/14/20002. This research has made use of the SIMBAD database, operated at CDS, Strasbourg, France.

\section*{Availability of Data}
The data underlying this article will be shared on reasonable request to the corresponding author.




\bibliographystyle{mnras}
\bibliography{SSW-bibtex} 





\bsp	
\label{lastpage}
\end{document}